# Electrochemical tuning of alcohol oxidase and dehydrogenase catalysis *via* biosensing towards butanol-1 detection in fermentation media

## Authors

Daria Semenova[a,*], , Tiago Pinto[a], Marcus Koch[b], Krist V. Gernaey[a] and Helena Junicke[a]

a Process and Systems Engineering Center (PROSYS), Department of Chemical and Biochemical Engineering, Technical University of Denmark, Søltofts Plads, Building 228A, 2800 Kgs. Lyngby, Denmark

b INM – Leibniz Institute for New Materials, Campus D2 2, 66123 Saarbrücken, Germany

*e-mail: dsem@kt.dtu.dk

## Abstract

A novel approach for electrochemical tuning of alcohol oxidase (AOx) and alcohol dehydrogenase (ADH) biocatalysis towards butanol-1 oxidation by incorporating enzymes in various designs of amperometric biosensors is presented. The biosensors were developed by using commercial graphene oxide-based screen-printed electrodes and varying enzyme producing strains, encapsulation approaches (layer-by-layer (LbL) or one-step electrodeposition (EcD)), layers composition and structure, operating conditions (applied potential values) and introducing mediators (Meldola Blue and Prussian Blue) or Pd-nanoparticles (Pd-NPs). Simultaneous analysis/screening of multiple crucial system parameters during the enzyme engineering process allowed to identify within a period of one month that four out of twelve proposed designs demonstrated a good signal reproducibility and linear response (up to 14.6 mM of butanol) under very low applied potentials (from -0.02 to -0.32 V). Their mechanical stability was thoroughly investigated by multi-analytical techniques prior to butanol determination in cell-free samples from an anaerobic butanol fermentation. The EcD-based biosensor that incorporates ADH, NAD$^+$, Pd-NPs and Nafion showed no loss of enzyme activity after preparation and demonstrated capabilities towards low potential (-0.12 V) detection of butanol-1 in fermentation medium (4 mM) containing multiple electroactive species with almost 15 times enhanced sensitivity (0.2282 µA/mM ± 0.05) when compared to the LbL design.

Furthermore, the ADH-Nafion bonding for the *S. cerevisiae* strain was confirmed to be 3 times higher than for *E. coli*.

## 1. Introduction

Being widely used as a chemical precursor in polymers and plastics production, butanol has gained a lot of attraction in the past years as potential replacement for petrochemical derived liquid fuels (Amiri and Karimi, 2019; Ibrahim et al., 2018). In contrast to ethanol, the most commonly used liquid biofuel, butanol possesses a higher energy density, hydrophobicity, boiling and flash points and can be blended with gasoline in any ratio without requiring modifications to current combustion engines (Amiri and Karimi, 2019). Furthermore, biobutanol production is well-established at industrial scale through a two-stage acetone-butanol-ethanol (ABE) fermentation process using solvent-producing bacteria (Amiri and Karimi, 2019). First, cellular growth occurs along with acetic and butyric acid production during the acidogenesis stage. This step is followed by solventogenesis, where the acids are re-assimilated into acetone, butanol, and ethanol, usually in a 3:6:1 ratio, respectively (Green, 2011). As solvent-producing bacteria, *Clostridium* species, viz. C. acetobutylicum, C. beijerinckii, *C. saccharoperbutylacetonicum* and *C. saccharobutylicum*, are the most extensively studied for butanol production (Jones and Woods, 1986; Wang et al., 2017). Due to high amylolytic activities, *Clostridium* species can utilize either simple or complex substrates, ranging from pure glucose to hydrolysates of corn and rice (Kiyoshi et al., 2015), microalgae (Wang et al., 2017), or palm oil (Al-Shorgani et al., 2015). While butanol remains one of the dominant products, it inhibits the growth of solvent-producing bacteria when present in excess (Chauvatcharin et al., 1995). Furthermore, the presence of many other compounds together with cells in the fermentation broth eventually makes the off-line monitoring and control of such processes using common analytical approaches, such as chromatographic and/or spectroscopic detection techniques, a rather time and resource consuming procedure (Pontius et al., 2020). Therefore, the design of an accurate and targeted analytical platform towards fast butanol detection and control is strongly required.

The alternative solution can be found in the development of an electrochemical biosensor sensitive to butanol. Driven by accurate and diverse point-of-care diagnostics within the biosensors research, a great variety of novel bioreceptors (i.e. enzymes, antibodies, etc.) combined with advanced electrical interfaces (i.e. nanoparticles, electrodes, etc.) were screened and validated in the past. On the one hand, such extensive biosensors development has broadened their application towards mapping and monitoring of various targets such as whole bacterial cells (e.g. *Escherichia coli*, *Salmonella typhimurium*) (Ahmed et al., 2014), viruses (e.g. HIV, hepatitis B) (Saylan et al., 2019), individual metabolites (e.g. glucose, ethanol) and electrolytes (e.g. sodium, chloride ions) (Kim et al., 2019). On the other hand, being mainly focused on the design of environmental and health monitoring tools, the biosensor detection limits of interest were reduced to the micromolar and low millimolar concentration range (e.g. for enzymatic ethanol sensitive biosensors the linear response can be rarely obtained above 30 mM). Nevertheless, the biosensor design was significantly improved towards fast and accurate analyte detection in biosamples (i.e. soil, blood, marine sediment, sweat, etc.), as well as to avoid biofouling (Kim et al., 2019). Not surprisingly, the application of such devices towards monitoring and control of fermentation processes (Cañete et al., 2018; Pontius et al., 2020) has gained considerable attention.

However, until now the potential of biosensors in the biotechnology production field could not be fully exploited due to the limited number of applications, as well as the limited number of species being currently monitored (mainly glucose and ethanol). For instance, the second most studied biosensor after glucose for fermentation applications is based on ethanol detection and has been successfully applied in wine production monitoring (Vasilescu et al., 2019). The main concept behind designing selective alcohol biosensing systems is based on enzyme immobilization, namely alcohol oxidase (AOx) or alcohol dehydrogenase (ADH), over various transducer surfaces (Pt, Au, glassy carbon, etc.) and applying mainly voltammetric or chronoamperometric detection methods (Alpat and Telefoncu, 2010; Azevedo et al., 2005; Lansdorp et al., 2019) in the presence of the analyte. Furthermore, the introduction of a mediator (e.g. conducting organic salts, quinone derivatives, etc.) either inside the biosensor system (Mechanism II, Figure 1) or in the analyte solution, allows faster electron shuttling from the enzyme redox active site to the electrode surface which subsequently prolongs the biosensor lifetime and enhances its sensitivity (Alpat and Telefoncu, 2010). Due to the high sensitivity and selectivity of AOx and ADH, the use of enzymatic biosensors towards low-carbon alcohol detection was shown to be a reliable, fast and relatively cheap solution (Hooda et al., 2018). The operational principles of AOx-based biosensors either rely on $H_2O_2$ formation or $O_2$ consumption (Mechanism I, Figure 1), whereas the operation of ADH-based biosensors is dependent on the direct oxidation of alcohols by means of nicotinamide cofactor $NAD^+$ present in the system (Mechanism III, Figure 1). Although a wide variety of research related to alcohol biosensing using enzymes can be found, a thorough analysis of the effect of enzyme producing strains, alcohol chain structure, pH, temperature, operational conditions, electrode and membrane materials on the electrochemical response of proposed biosensor designs is still missing. Thus, by varying experimental conditions (pH, enzyme load, different alcohols, temperature) Patel et al. (Patel et al., 2001) studied the response of AOx-based sensors from different strains (*Hansenula sp.*, *Candida boidinii*, *Pichia pastoris*) and demonstrated their low catalytic activity in the presence of propanol-1 and butanol-1 (compared to ethanol and methanol < 5 %). Guilbault et al. (Guilbault et al., 1983; Nanjo and Guilbault, 1975) performed the analysis of multiple biosensing system parameters including the variation of the enzyme producing strains sensitivity and selectivity towards detection of various primary alcohols and demonstrated how the sensitivity of the system can be completely reversed after enzyme immobilization (when compared to free enzyme solution) over the electrode. Furthermore, an additional presence of a hydrophobic membrane can eliminate the pH dependency of the enzyme-based biosensors (Lubrano et al., 1991). In the comparison Table S1 (Supplementary Material) it was shown for alcohol biosensors that a choice of the right combination of enzyme source and type with a proper electrical interface, membranes, operational conditions and fabrication methods may allow to optimize their catalytic activity (as well as linear detection limits) towards given substrates and reaction environments. In other words, electrochemical tuning *via* biosensing can improve the enzymatic activity towards specific detection of the desired analyte (Devine et al., 2018; Saboe et al., 2017) and as a result enable engineering of biocatalysts (as a part of the biosensing system) to fit the required bioprocess.

The biosensor development reported in this paper has been part of a larger research effort aiming at establishing measurement techniques for monitoring continuous fermentation based production of butanol from waste streams with mixed cultures that convert butyrate in the presence of hydrogen. Here, we report for the first time the development of a butanol sensitive amperometric biosensor. The behaviour of commercially available AOx and ADH enzymes was studied in 12 different biosensor designs towards butanol-1 detection in liquid media. Furthermore, the role of fabrication approaches, including layer-by-layer immobilization, one-step electrodeposition together with enzymes, membrane films and Pd-nanoparticles (Pd-NPs), and the use of mediators (Meldola Blue and Prussian Blue) in the enhancement of the biocatalytic performance were investigated. Finally, the analytical merit of the most promising biosensor designs was evaluated in the presence of butanol-1 containing fermentation samples.

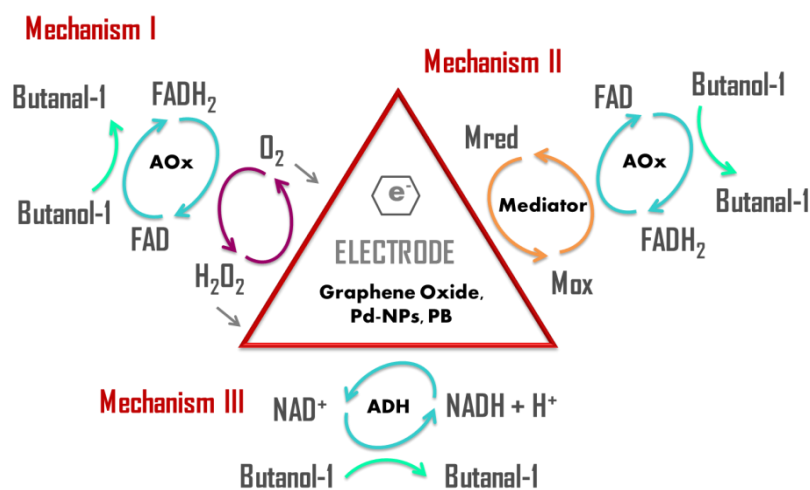

Figure 1 – Summary of butanol-1 oxidation mechanisms evaluated in the presented study for alcohol oxidase (AOx) and alcohol dehydrogenase (ADH) based biosensors.

## 2. Materials and methods
### 2.1. Instrumentation and reagents

Alcohol oxidase (AOx) (EC 1.1.3.13, from *Pichia pastoris*, 32 U/mg solid), alcohol dehydrogenase (ADH) (EC 1.1.1.1; from *Saccharomyces cerevisiae*, 321 U/g solid), β-Nicotinamide adenine dinucleotide sodium salt (NAD$^+$) and bovine serum albumin (BSA) (lyophilized powder, ≥96.0%) were obtained from Sigma (St. Louis, MO, USA). ADH-175 (lyophilized powder, 24.5 U/mg) from recombinant *Escherichia coli* strain was purchased from c-LEcta GmbH (Leipzig, Germany). Note that ADH obtained from *S. cerevisiae* and *E. coli* will be further referred in this paper as ADHs and

ADHc, respectively. Glutaraldehyde solution (25% (v/v)), methanol (UV HPLC gradient, 99.9%), ethanol (UV HPLC gradient, ≥99.9%), butanol (UV HPLC gradient, ≥99.7%), Meldola Blue and potassium hexacyanoferrate (III) (ACS reagent, ≥99.0%) were purchased from Sigma – Aldrich (St. Louis, MO, USA). Iron (III) chloride (anhydrous, 99.99%) and Nafion®117 solution (~5% (v/v) in a mixture of lower aliphatic alcohols and water) were provided by Aldrich (Steinheim, Germany). Mono – and di-potassium hydrogen phosphates (anhydrous) were obtained from Merck (Darmstadt, Germany). Sodium hydroxide (50% (w/w)) and hydrochloric acid (37% (w/w)) solutions were purchased from VWR International A/S (Søborg, Denmark). All the solutions were prepared with 0.1 M phosphate buffer supplemented with 0.1 M KCl (PBS, pH = 6), unless stated otherwise.

For preparation of electroplated nanoparticles, the following reagents were utilized: $H_2PdCl_4$, $(NH_4)_2HPO_4$ and $Na_2HPO_4 \cdot 12H_2O$, 99 % purity (Sigma Aldrich). The $Pd^{2+}$ standards for ICP-MS measurements were obtained from Honeywell Fluka (Fisher Scientific, Germany).

For biosensor preparation, DRP-110DGPHOX (DropSens, Llanera, Spain) screen printed electrodes (SPEs) printed on ceramic substrate were used. Each sensor consists of a carbon working electrode (WE) (0.4 cm diameter; 0.126 $cm^2$ apparent geometric area) modified with a graphene oxide (GO) layer, a carbon counter electrode and a silver (Ag/AgCl) reference electrode. Biosensors electrochemical characterization (cyclic voltammetry (CV) and chronoamperometric (AM) measurements) were performed using MultiEmStat (PalmSens, Utrecht, The Netherlands) with a DRP-CAST1X8 interface (DropSens, Llanera, Spain) controlled by MultiTrace 3.4 software (PalmSens, Utrecht, The Netherlands).

Dissolved oxygen (DO) concentrations either in the AOx stock solutions or in the analyte droplet for AOx-based biosensors were monitored using an OXR430 retractable needle-type fiber-optic oxygen minisensor (PyroScience GmbH, Aachen, Germany). The sensor was connected to a FireStingO2 fiber-optic meter (PyroScience GmbH, Aachen, Germany) and DO consumption rates were monitored by Pyro Oxygen Logger software (PyroScience GmbH, Aachen, Germany).

## 2.2. Biosensor preparation and characterization

In order to investigate the ability of the biosensor's design (i.e. layer composition, sequence, etc.) to tune the biocatalytic activity of enzymes, two main preparation approaches were proposed: electrochemical co-deposition (EcD) and layer-by-layer (LbL) enzyme deposition (Figure S1, Supplementary Material). Following the procedure described in (Semenova et al., 2020), enzymes were encapsulated and co-deposited in one step together with Pd-NPs and with/without Nafion®117. The Pd-based polyelectrolyte was prepared by mixing $H_2PdCl_4$ (3.5 g/L), $(NH_4)_2HPO_4$ (20.0 g/L), and $Na_2HPO_4 \cdot 12H_2O$ (100 g/L) solutions. In case of AOx-based biosensors, a 4 μL droplet prepared, with an equal volume ratio of AOx (0.034 U/μL), BSA (5 vol% diluted in water) and Pd-NPs mixture with or without Nafion (2% vol., neutralized in water) was dropped over a GO-modified WE and a 2.5 mA current was applied for 30 sec through a Pd reference wire immersed into the droplet. For ADH-based biosensors, a 4 μL mixture, containing ADH (for both, ADHs and

ADHc based designs, 1 mg of enzyme was dissolved in 1 mL of PBS), NAD$^+$ (83 mM in PBS), Pd-NPs and Nafion (2% vol., neutralized in water) in a 1:1:1:1 proportion (v/v/v/v) was co-deposited at the same electroplating conditions. For AOx-based LbL designs, a 4 μL droplet of enzyme solution, mixed in a 1:1 proportion (v/v) with BSA (5 vol% diluted in water), was cross-linked using glutaraldehyde (2.5 μL, 1% (v/v) diluted in water) to the WE surface, modified either with 0.1 M solution of Prussian Blue or Meldola Blue, and then covered with Nafion film (4 μL droplet of 2% vol., neutralized in water). The Prussian Blue (PB) deposition and LbL biosensor preparation approaches were adopted from (Semenova et al., 2019). The ADH-based biosensors were prepared by direct placement of enzyme/BSA mixture (1:1 proportion (v/v)) over the WE surface modified with glutaraldehyde, followed by stepwise immobilization of NAD$^+$ and Nafion layers. As another approach, ADH was immobilized over the WE surface modified with NAD$^+$ and then covered step-by-step with individual BSA, glutaraldehyde and Nafion films. In case of Pd-NPs based LbL biosensors, the NPs were electrodeposited over the WE surface prior to any layer formation. In order to guarantee the uniform layer formation of LbL designs, the biosensors with freshly deposited solution (enzyme, BSA, etc.) were placed at least for 8 hours to dry in a climate chamber at 40% of humidity and 8 °C. Regardless the fabrication method, SPEs were not pre-treated prior to the modification steps (Semenova et al., 2019) and at the final preparation step, butanol biosensors were washed with deionized (DI) water, air dried and stored before use in the dark at 4° C. All the designs of potentially butanol sensitive biosensors summarized in Table 1 were constructed to minimize the applied potential values in order to eliminate the occurrence of side reactions at high potentials and to prevent high irreversibility NAD$^+$/NADH redox reaction (for ADH-based designs), as well as undesired oxygen reduction at high positive potentials (for AOx-based designs). Note that PB and Pd-NPs were also introduced to facilitate the hydrogen peroxide decomposition (Semenova et al., 2020, 2018) for AOx-based biosensor designs and thus improve the enzyme activity and overall operational stability of the system. Furthermore, the presence of MB and Pd-NPs was supposed to improve the electron transfer between the enzyme-containing layer and electrode surface (Santos et al., 2003; Semenova et al., 2020).

The CVs were obtained by placing a 100 μL droplet of PBS over all three electrodes and applied 20 mV/sec scanning speed (from −0.5 V to +0.5 V potential range, unless stated otherwise). The results of GO modification for all proposed designs were summarized in Figure 2 and Supplementary Material (Figure S2). Based on obtained peak potentials in the CVs, various applied potential values were verified during AM characterization of butanol-1 (concentration range: 0 -14.6 mM). All measurements from the same biosensor design were repeated at least in triplicates.

Table 1 – Summary of enzymatic alcohol biosensor designs and applied current values ($E_{app}$) evaluated towards butanol-1 detection. The important biosensing components, such as palladium nanoparticles (Pd), alcohol oxidase (AOx), alcohol dehydrogenase (ADH), Nafion (Naf), Prussian Blue (PB), Meldola Blue (MB), nicotinamide adenine dinucleotide (NAD) and glutaraldehyde (Glut), were highlighted for each design.

| Enzyme deposition method | Name | Composition | $E_{app}$, V |
|---|---|---|---|
| EcD | design 1 | Pd/AOx | -0.23 |
| | design 2 | Pd/AOx/Naf | -0.32<br>-0.02 |
| | design 3 | Pd/ADH$_c$/NAD/Naf | -0.12 |
| | design 4 | Pd/ADH$_s$/NAD/Naf | -0.12 |
| LbL | design 5 | PB/AOx/NaF | -0.23 |
| | design 6 | MB/AOx/Naf | -0.23 |
| | design 7 | Pd/AOX/Naf | -0.32<br>-0.02 |
| | design 8 | Pd/PB/AOX/Naf | 0.02<br>-0.02 |
| | design 9 | NAD/ADHs/BSA/Glut/Naf | 0.25<br>0.05<br>0.00<br>-0.10<br>-0.20<br>-0.35<br>-0.32 |
| | design 10 | ADHs/NAD/Naf | -0.32 |
| | design 11 | ADHc/NAD/Naf | |
| | design 12 | Pd/ADHs/NAD/Naf | -0.32 |

### 2.3. Butanol fermentation setup and conditions

Anaerobic granular sludge from an industrial effluent treatment BIOPAQ®IC (Novozymes, Kalundborg, Denmark) reactor was selected as inoculum. The original sludge is used for the treatment of a feed containing, among others, organic acids, making it an ideal candidate for butyrate fermentation. This assumption was previously tested and validated (data not shown) by the authors, to a successful production of butanol via butyrate fermentation in the presence of hydrogen. In this work, butanol production was achieved in a bioreactor with 2 L working volume (Applikon, Delft, The Netherlands) inoculated with 10 % (v/v) sludge and operated under anaerobic conditions by continuous sparging with molecular hydrogen (0.050 L$_N$/min). The temperature in the reactor liquid was maintained at 37 °C and the pH at 6.4±0.1 (AppliSens, Applikon Biotechnology, The Netherlands) through addition of 2 M NaOH and 2 M HCl. Temperature, pH, and stirring speed (400 rpm) were controlled by an ez-Control system (Applikon, Delft, The Netherlands). The inflow of hydrogen gas was controlled with a red-y smart controller GSC (Vögtlin, Aesch, Switzerland). The fermentation medium contained 3 g/L yeast extract, 10 g/L butyric acid, 2.2 g/L acetate, 0.5 g/L $KH_2PO_4$, 0.5 g/L $K_2HPO_4$, 1 g/L NaCl, 0.2 g/L $MgSO_4·7H_2O$, 0.01 g/L $MnSO_4·H_2O$, 0.01 g/L $FeSO_4·7H_2O$, 0.1 g/L thiamin, 0.001 g/L biotin. All chemicals were acquired from Sigma-Aldrich (St. Louis, Missouri, USA). Sampling was conducted at pre-determined time intervals, aseptically and directly from the reactor fermentation medium. Samples were centrifuged at 13000 rpm (MiniSpin Plus, Eppendorf, Germany) and then filtrated through a 0.22 μm pore size filter (Minisart® Syringe filter, Sartorius, Germany) before storage at 4 °C until further analysis.

### 2.4. Enzyme preparations and stock activity measurements

The immobilization of enzymes together with Pd-NPs over the sensors surfaces did not affect their activity when compared to the activity of stock solutions, as has been previously shown in (Semenova et al., 2020).

### 2.5. Scanning electron microscopy (SEM) and energy dispersive X-ray spectroscopy (EDX)

For morphological characterization and elemental analysis, a scanning electron microscope (FEI Quanta 400 FEG) equipped with an X-Ray spectrometer (EDAX Genesis V 6.04) was used. Prior to the measurements, the samples were dried under ambient conditions and investigated without further preparation. Both, secondary and back scattered electron images and X-Ray spectra were acquired in high vacuum mode at 10 kV accelerating voltage.

### 2.6. Liquid chromatography-electrospray ionization-tandem mass spectrometry (LC-ESI-MS/MS) or Q-TOF

To estimate the long-term stability of the nanobiosensors with encapsulated enzymes without co-elution of both enzyme and membrane, liquid chromatography-electrospray ionization coupled mass spectrometry (LC-ESI-MS) studies were conducted according to the earlier reported protocols (Semenova et al., 2019, 2018). All the experiments were performed on a quadrupole time-of-flight mass spectrometer Q-TOF LC/MS 6545 (Agilent Technologies, CA, USA) equipped with Jet Stream Thermal Focusing Technology ESI source. The data acquisition was controlled by the MassHunter Software Tool.

### 2.7. Inductively coupled plasma mass spectrometry (ICP-MS)

To verify the long-term mechanical stability of Pd-NPs onto GO-surface after one- or two-step co-deposition, ICP-MS analysis of BuOH containing probes collected after electrochemical calibration was performed on an ELEMENT XR (Thermo Fisher Scientific, Bremen, Germany) coupled with an auto sampler SC-E2 DX (Elemental Scientific, Omaha, USA). For the analysis, $Pd^{105}$ isotope (magnet set at m/z 104.9) was measured at high resolution mode with the following source parameters: cool gas, 16.00 L/min; sample gas, 1.160 L/min; Faraday deflection, -218 V; plasma power, 1250 W; peristaltic pump speed, 10 rpm; torch X-Pos., 2.1 mm; torch Y-Pos., 0.9 mm; torch Z-Pos., -4.0 mm. The detector was set at 1700 V. All calibration solutions and blank samples were supplemented with 2% HCl prior to ICP-MS analysis. All measurements from the same batch were performed at least in triplicate and the final results were expressed as mean values with relative standard deviation.

### 2.8. Gas chromatography mass spectrometry (GC-MS) assays for analysis of real fermentation samples containing butanol-1

For BuOH quantification in the presence of phosphate buffer, the GC-MS assay was optimized. For the experiments, a GC-MS system equipped with a WAX-plus column (Torrance, CA, USA) (30 m×0.25 mm; film thickness 0.25 µm) was used. The injection temperature was set at 200°C (split ratio 1:30) and the injection volume was 1 µl. The column temperature was set at 200 °C for 5 min. All calibration solutions and blank samples were prepared in phosphate buffer, pH=7. All measurements from the same batch were performed at least in triplicate.

For separation and analysis of fermentation media depending on cultivation time, the following temperature program was set starting from 60 °C for 1 min, then raised to 200 °C at 20 K/min and held at the final temperature of 250 °C for 5 min, (split ratio 1:250). The transfer line to the mass spectrometer and the source temperatures were 220°C and 200°C, respectively. The ionization of the compounds was performed at an acceleration voltage of 70 eV. Mass spectra were recorded in TIC and EIC modes at the m/z range of 40-600.

## 3. Results and discussion
### 3.1. Biosensor design tuning towards butanol detection

Previously we have shown that by optimizing fabrication approaches, operating conditions, composition and design of individual biosensor layers, the enzyme activity towards the desired analyte can be significantly improved, as well as the overall biosensor performance, lifetime and detection range (Semenova et al., 2020, 2019, 2018). Although AOx and ADH activities rapidly decrease in the presence of alcohols with a carbon number greater than two, it was interesting to investigate how incorporating the enzyme as a part of the biosensing system can potentially improve its biocatalytic properties towards butanol oxidation. Therefore, the analytical merit of biosensor designs possibly sensitive towards butanol-1 in Table 1 was investigated during chronoamperometric studies. Furthermore, knowing that the biosensor response is strongly dependent on the applied potential ($E_{app}$) (Santos et al., 2003), based on obtained voltammograms (Figure 2A-C), various $E_{app}$ values were evaluated. As a result of electrochemical characterization, four types of biosensors, namely *design 2*, *design 4*, *design 8*, and *design 11*, demonstrated good signal reproducibility and linear response within a 0 to 14.6 mM butanol-1 concentration range (Figure 2D) under very low applied potential values (-0.02 V for *design 2* and *design 8*; -0.12 V for *design 4*; and -0.32 V for *design 11*). Hence, AOx co-deposited together with Pd-NPs in *design 2*, demonstrated a relatively broad detection range (up to 8 mM BuOH) and an almost 3 times higher sensitivity (0.0873 µA/mM ± 0.006) than for LbL *design 8*, where the additional presence of the PB mediator together with Pd-NPs facilitated $H_2O_2$ reduction during butanol conversion that eventually resulted in an extended linear detection range up to 14.6 mM of BuOH. As for ADH-based biosensors, although both *design 4* and *design 11* worked up fine to 4 mM BuOH, enzyme co-deposition together with Pd-NPs significantly enhanced (almost 15 times higher) the *design 4* biosensor sensitivity (0.2282 µA/mM ± 0.05). Remarkably, the EcD encapsulation approach did not decrease neither ADH nor AOx activity towards butanol-1 oxidation when compared to the activity of free enzyme solutions (data not shown), similar to the previously obtained results (Semenova et al., 2020). Nevertheless, in order to understand the nature of such variation in biosensor performance, the mechanical stability of the proposed designs was studied in section 3.2.

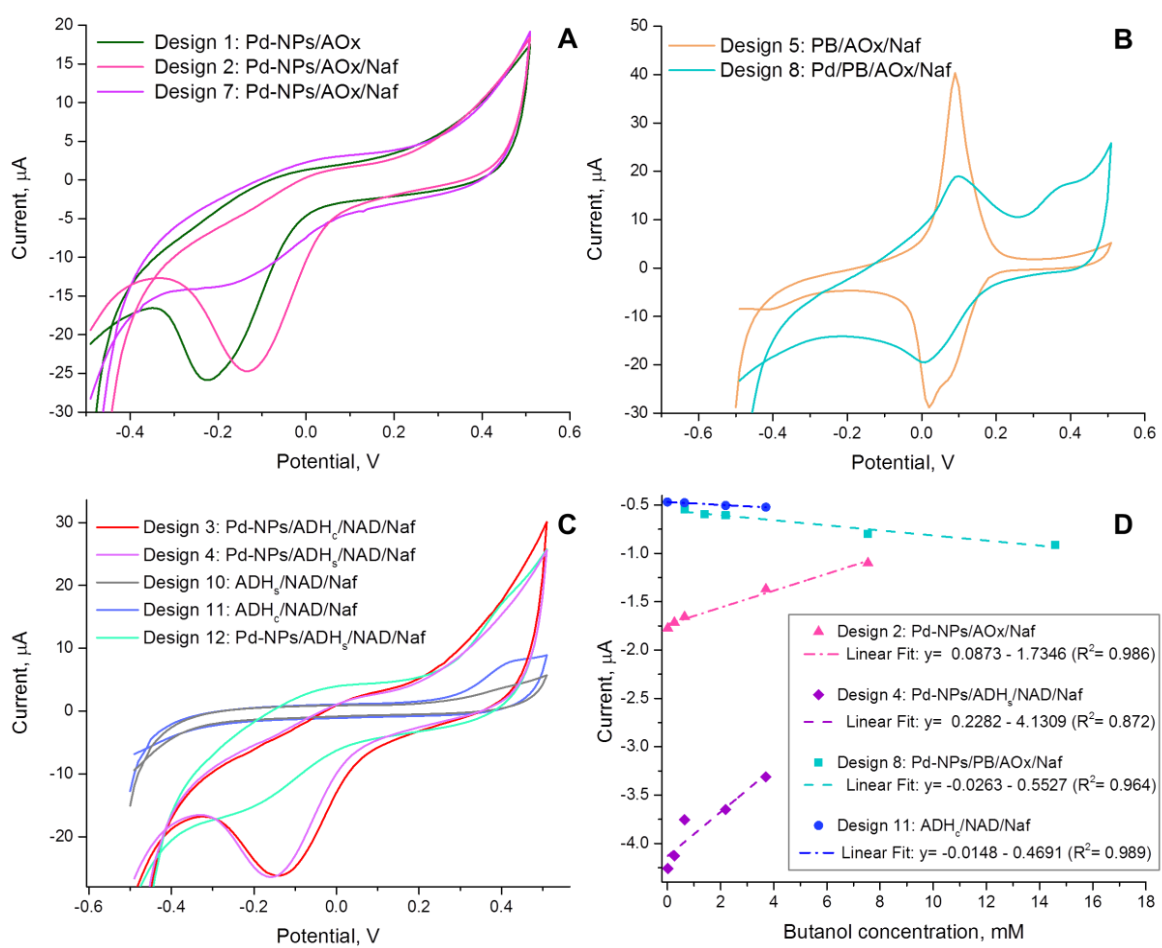

Figure 2 – Comparison of cyclic voltammetric (A-C) and chronoamperometric (D) responses of AOx- and ADH- based butanol-1 biosensors. Voltammograms were obtained in air saturated PBS (pH = 6) at 20 mV/sec scanning speed for various biosensors. In *design 1 – design 4*, the enzyme was co-deposited together with Pd-NPs and with/without Nafion (Naf) NAD$^+$ co-factor (required for ADH-based designs). The rest of the designs (*design 5 –design 12*) correspond to the layer-by-layer fabrication approach, where enzymes were either placed over the bare working electrode surface ormodified with Pd-NPs or Pd-NPs covered by a Prussian Blue layer.For ADH-based designs the immobilized enzyme was further covered by NAD$^+$ and Nafion solution neutralized in water as a final covering layer. The calibration curves in D were registered at -0.02 V (*design 2* and *design 8*), -0.12 V (*design 4*) and -0.32 V (*design 11*). The butanol concentrations ranged from 0 to 14.6 mM. The relative standard deviation per biosensor is not shown due to the low value (RSD = 9%).

### 3.2. Analysis of mechanical stability after performing calibrations with butanol-1

By knowing that co-deposition of Pd-NPs with Nafion and enzymes over the WE of biosensors results in the formation of uniform and stable biofilms (Semenova et al., 2020), further investigation of the mechanical stability of biosensors fabricated by the EcD approach *vs* LbL method during electrochemical characterization with analyte (butanol) was required. In order to reduce the influence of lower aliphatic alcohols present in a commercial Nafion®117 solution on the enzyme (both AOx and ADH) activity and/or stability during the fabrication step, Nafion neutralized in water was applied for all designs presented in Table 1. Hence, it allowed to obtain visually uniform layer formation for all designs and additionally for LbL biosensors to avoid numerous damages and structural defects, previously discussed in (Semenova et al., 2020). Furthermore, a significant role of Nafion in transport phenomena between EcD and LbL designs can clearly be seen already at the CV characterization step with buffer probes (Figure 2A-C). More precisely, *design 2* and *design 7* are having the same composition (Pd/AOx/Naf) but when comparing their voltammograms (Figure 2 A) with pure Pd-NPs deposited over GO (Figure S2, Supplementary Material) or with EcD *design 1* prepared without Nafion, no characteristic Pd-NPs peak potential was obtained for the LbL fabricated biosensors. A similar Pd-NPs peak hindrance was obtained for Lbl *design 12* of the ADH- based biosensor when compared with corresponding EcD *design 4* (Figure 2C). Therefore, based on the obtained voltammograms, it was decided to investigate the response of the biosensors in diffusion and kinetic control regions and vary the applied potential values similar to (Boujtita et al., 2000).

In Figure 3 the results of combined SEM/EDX investigation of various biosensor layers during fabrication and after chronoamperometric studies with butanol-1 were summarized. The direct electroplating of Pd-NPs from (poly)electrolyte for 30 sec resulted in homogenous deposition of individual spherical particles (Figure 3A) or hybrid organic–inorganic nanoparticles containing enzyme, Nafion and Pd-NPs for EcD designs (Semenova et al., 2020). Moreover, individual Pd-NPs allowed to cover the cavities present in GO (Semenova et al., 2018), which subsequently allowed further uniform deposition of the Prussian Blue layer (Figure 3B), AOx, etc. for LbL designs. From Figure 3D and 3E, no structural defects in SEM images for *design 2* (EcD) and *design 11* (LbL), respectively were captured after performing the calibration with pure butanol-1 solutions, as well as no complete loss of deposited components in EDX spectra (Figure 3F) was obtained. A similar behaviour was confirmed for *design 4* and *design 8* biosensors (data not shown).

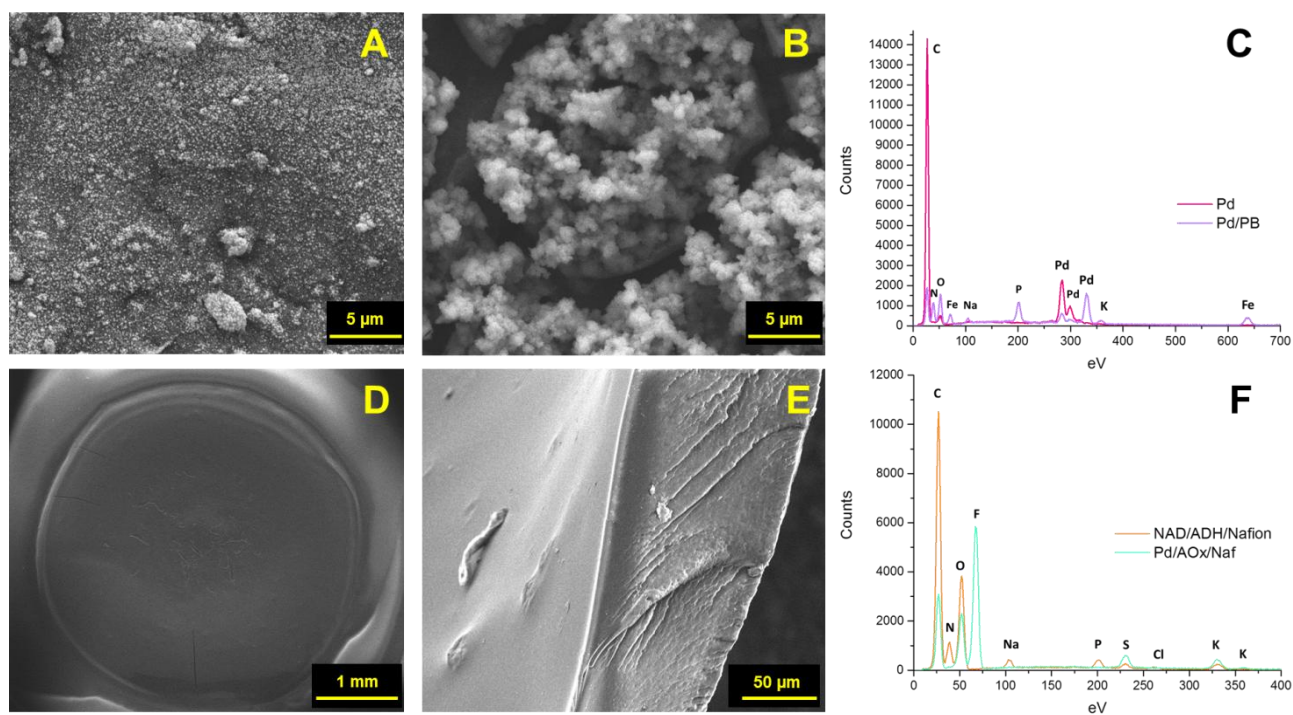

Figure 3 – SEM images of Pd-NPs (A) electroplated for 30 sec over the working electrode and Pd-NPs covered with Prussian Blue (Pd/PB) particles (B) with corresponding EDX spectrum shown in (C). The uniformity of deposited layers for Pd/AOx/Nafion (*design 2*, top view) and NAD/ADH/Nafion (*design 11,* cross-section) biosensors after performed voltammetric and chronoamperometric characterization is demonstrated in the SEM images (D) and (E), respectively. The corresponding EDX spectrum is presented in (F).

However, by investigating $Pd^{105}$ and Nafion elution rates after performing AM calibration with butanol-1, the difference in biosensor stability and response got more pronounced for various designs. As expected, Pd leakage was significantly lower for EcD-based designs when compared to LbL analogues, as shown in Figure 4A, with one exception (*design 3* with ADH from *E. coli* strain). Interestingly, the Nafion elution rate for biosensors containing ADH from *E. coli* (*design 11*) was almost 3 times higher when compared to ADH from *S. cerevisiae (design 10)*. We assume these phenomena can be in direct relation with variations in structural stability and conformation of ADH enzymes obtained from different strains that eventually affects its bonding either when co-deposited with Pd-NPs and Nafion or directly attached to a GO surface. Therefore, from Figure 4 biosensors based on ADH from *S. cerevisiae* (i.e. *design 4, design 11*) yield a more robust design regardless the fabrication approach which subsequently prolongs their operational stability. Note that due to the fully instrumentally controlled fabrication process EcD-based enzymatic biosensors have a significant advantage when compared to LbL analogues since it allows in a simplified manner to modify the biosensor composition, as well as to guarantee sensor-to-sensor fabrication and measurement reproducibility (Semenova et al., 2020). All four selected designs underwent further evaluation with butanol fermentation media.

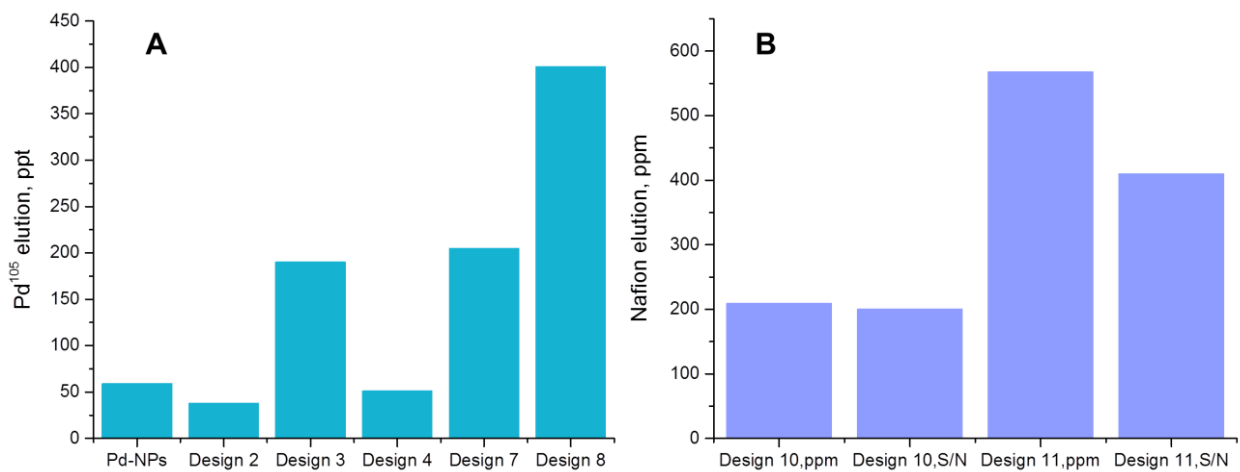

Figure 4 – The impact of electrochemical characterization of biosensors in the presence of butanol on their mechanical stability. The results of ICP-MS monitoring of palladium ($Pd^{105}$) elution from Pd-NPs monolayer electroplated over the working electrode and various butanol biosensor designs are summarized in (A). For ADH-based biosensors (*design 10* and *design 11*) probes were collected after the final chronoamperometric characterization with butanol in buffer and analysed for Nafion leakage *via* liquid chromatography-electrospray ionization-tandem mass spectrometry. The Nafion elution was identified in mass spectra due to the presence of the $CF_3(CF_2)_5CHFCOO-$ carboxylic acid fragment (peak area at m/z = 394.9758) and summarized for both designs with corresponding signal-to-noise (S/N) ration in (B).*Note:* RSD values are too low (less than 3%) to be shown at trace concentration levels from nanobiosensors (N=5).

### 3.3. Evaluation of biosensor response with real samples from anaerobic biobutanol production

A thorough biosensor development process requires a detailed characterization of selectivity, response time, testing volume and many other critical parameters (Kulkarni and Slaughter, 2016), crucial for efficient and robust functioning of such complex systems in various samples. From chromatograms in Figure 5 obtained after 21 days of anaerobic biobutanol production at lab-scale it is clearly seen that the presented medium consists not only of several electroactive species but also contains other aliphatic alcohols that potentially could compete with butanol detection when using biosensors. However, being resource-limited (mainly, restricted by the one month period of instrumental access to perform multi-analytical analysis) and aiming as a final step in butanol biosensor development its further application towards off-line monitoring of butanol-1 in real fermentations, an immediate evaluation of the previously proposed designs (see section 3.1.) with cell-free fermentation samples was performed. Since all four types of biosensors (*design 2, design 4, design 8 and design 11*) were able to detect variations in butanol concentrations up to 4 mM, the fermentation probe obtained after 288 hours and containing 2.21 mM of liquid BuOH (GC-MS analysis used as a reference method) was analysed by means of chronoamperometry. Hence, only the biosensor with *design 4*, where ADH was electrochemically co-deposited together with Pd-NPs and Nafion, was able to accurately detect butanol-1 (2.1978 mM ±0.05 from the calibration curve in Figure 2D) in the cell-free fermentation medium, as shown in Figure 5. For the remaining proposed the proposed designs, the AM current curves obtained for fermentation samples did not overlap with 2.19 mM AM curves of pure BuOH detected in buffer and have not reached a steady state within a 3 min time interval. It is known that alcohol biosensors based on ADH have a higher stability and specificity when compared to AOx (Azevedo et al., 2005), and additional EcD fabrication of ADH biosensors allows to avoid a decrease of the electron transfer which is common for highly cross-linked enzymes in LbL designs (Saboe et al., 2017). It is also important to mention that despite the need of external $NAD^+$ co-factor, a crucial advantage of ADH-based biosensors is that the sensing system is not dependent on the presence of oxygen, contrary to AOx-based biosensors, which does not restrict the areas of application and potentially opens up the possibilities for their further optimization and integration for on-line monitoring of anaerobic biobutanol production.

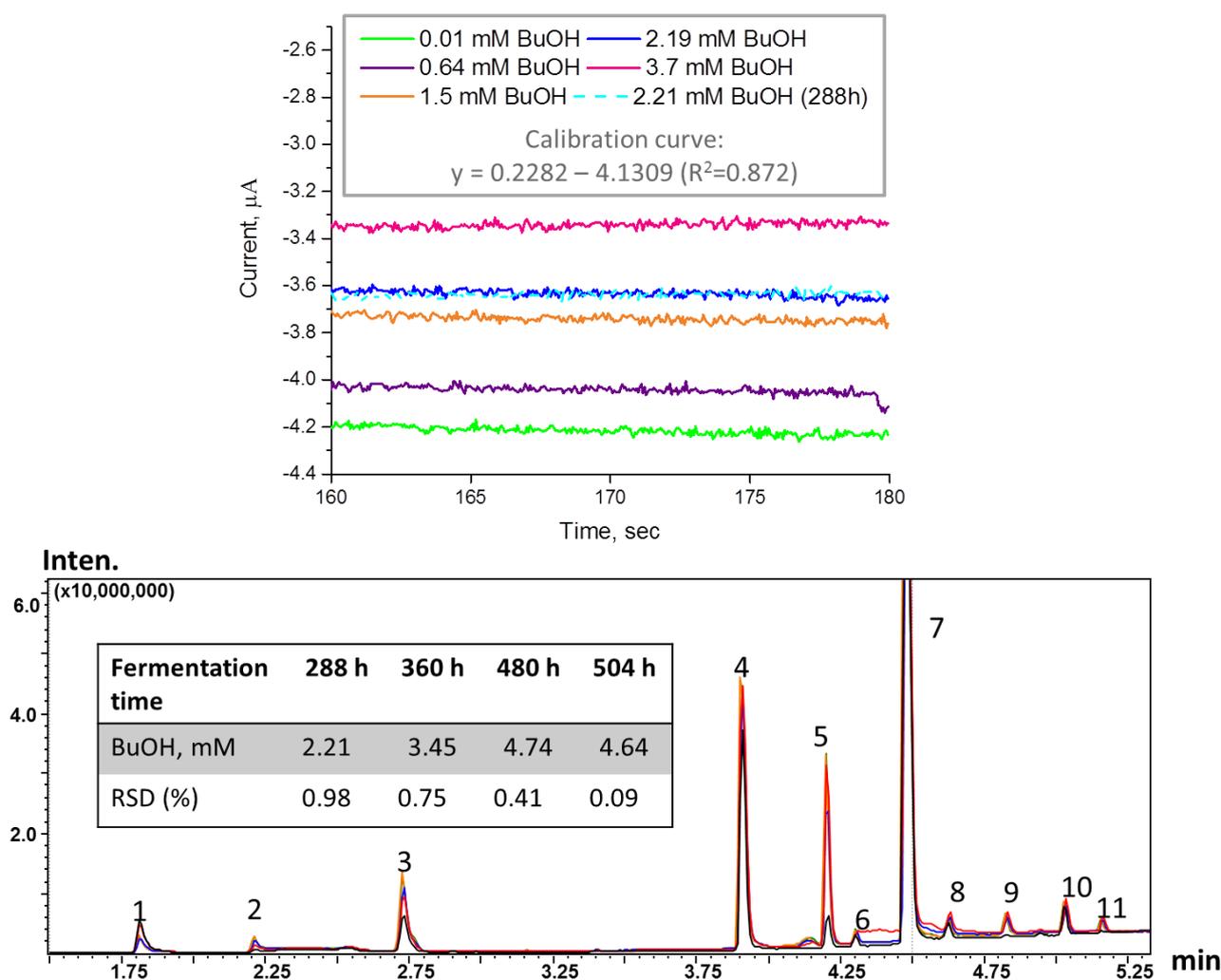

Figure 5 - Chronoamperometric response (current *vs* time) of one-step encapsulated ADH-based biosensor (*design 4*) recorded for pure butanol probes (0.01–3.7 mM) and cell-free sample containing 2.21 mM butanol after 288 h of fermentation (*top*). The calibration curve (shown in Figure 2) used for extrapolation of the butanol concentration during AM characterization of fermentation sample is highlighted. Chromatograms obtained by GC-MS analysis of fermentation samples (*bottom*) collected at different points in time: black – 288 h; red – 360 h; blue – 480 h; orange – 504 h. The butanol-1 concentration obtained in a liquid phase of each probe is summarized in the table. Note: (1) ethanol; (2) propanol-1; (3) butanol-1; (4) acetic acid; (5)-propionic acid; (6) 2-methyl-propanoic acid; (7) heptanoic acid; (8) buthanoic acid; (9) pentanoic acid; (10) acetamide; (11) butanamide.

## 4. Conclusions and future perspectives

By combining operating conditions, layers composition and deposition approaches, AOx and ADH biocatalytic behaviour was successfully tuned electrochemically towards butanol-1 detection under very low applied potential values in four different designs of amperometric biosensors. Furthermore, the obtained variation in response and linear detection limits of butanol sensitive biosensors was fully interpreted in terms of mechanical stability by means of multi-analytical techniques. Thus, the EcD fabrication approach demonstrated a significantly higher stability and sensitivity of biosensors in the presence of the analyte. Remarkably, it was shown for the first time that depending on the strain producing ADH, a significant difference in stability of enzyme-Nafion bonding was obtained which eventually affects not only mechanical properties, but also the overall performance and lifetime of the biosensing system. As a final step, the performance of butanol biosensors was evaluated off-line with cell-free butanol fermentation samples. The biosensor design based on electrochemical co-deposition of ADH together with Pd-NPs, $NAD^+$ and Nafion (*design 4*) demonstrated the capabilities towards butanol-1 detection in a complex matrix containing multiple electroactive species together with other low-carbon number alcohols present in the fermentation medium. The absence of oxygen limitations of *design 4* biosensor (due to the presence of ADH) together with improved sensor-to-sensor fabrication and measurement reproducibility (additional advantages of the EcD fabrication method) allows its further promotion towards development of a commercial tool for at- or on-line monitoring of butanol fermentations.

Another practical aspect of the presented work was to demonstrate that electrochemical tuning of enzymes being a part of the biosensing system allows to optimize the biocatalytic behaviour for a given substrate and environment in a gentle manner. Knowing that biosensors development together with evaluation of individual system parameters on electrochemical response is a rather time and resource consuming process , the presented enzyme engineering approach performed within a one month period significantly shortens the transition time between the desig stage and biosensor application for monitoring of the desired analyte in various media. It would be fair to say that the presented concept would be mostly suitable towards biosensor development for the analysis of fermentation samples rather than for healthcare applications, where the number and the volume of expensive biological samples (i.e. blood, saliva, urine, etc.) are significantly lower. However, the experimental evaluation of multiple biosensing system parameters for various designs combined with mathematical modeling techniques can allow identifying the crucial combinations that affect the biosensor response, sensitivity, selectivity, etc. in a more effective way and much shorter times.

## Acknowledgments

The authors would like to acknowledge the support obtained from the European Union's Horizon 2020 research and innovation program under the Marie Sklodowska-Curie grant agreement number 713683 (COFUNDfellowsDTU), from the Danish Council for Independent Research in the frame of the DFF FTP research project GREENLOGIC (grant agreement No. 7017-00175A), from the Novo Nordisk Fonden in the frame of the Fermentation-Based Biomanufacturing education initiative (grant agreement No. NNF17SA0031362). The authors would like to acknowledge Prof. L. Micheli for her inputs and support during revision process; Dr. Y. E. Silina for participation in the development of the original manuscript and the provided ICP-MS and GS-MS analysis; Dr. P. W. de Oliveira and Dr. D. Beckelmann at INM for providing access to their instrumentation.